\documentstyle[aps,floats,epsf]{revtex}
\tighten
\draft
\newcommand{\postscript}[2]{\setlength{\epsfxsize}{#2\hsize}
   \centerline{\epsfbox{#1}}}

\begin{document}

\twocolumn[\hsize\textwidth\columnwidth\hsize\csname
@twocolumnfalse\endcsname

\hfill$\vcenter{\hbox{\bf IUHET-435} \hbox{May 2001}}$  

\title{Higgs Sector Radiative Corrections and $s$-Channel Production} 
 
\author{M.~S. Berger}
\address{Physics Department, Indiana University, Bloomington, Indiana 47405}
\date{May 14, 2001}
\maketitle

\begin{abstract}
Higgs boson mass sum rules of supersymmetric models offer attractive 
targets for precision tests at future muon colliders. These sum rules involve
the gauge boson masses as well as the masses of the Higgs boson states which 
can be precisely measured in the $s$-channel production process at a muon 
collider. These measurements can sensitively probe radiative corrections to 
the Higgs boson masses as well as test for CP-violation and nonminimality 
of the Higgs sector.
\end{abstract}

\vspace{0.25cm}]
\narrowtext

\newcommand{\be}{\begin{equation}}
\newcommand{\ee}{\end{equation}}
\newcommand{\bea}{\begin{eqnarray}}
\newcommand{\eea}{\end{eqnarray}}


{\bf Introduction:\ } 
In recent years the electroweak precision measurements have 
played a large role 
in establishing the validity of the Standard Model as well as constraining the 
possibilities of new physics. In particular the precision measurements narrowed
the allowed values for the top quark mass, and the top quark mass was 
observed directly and its mass is consistent with radiative corrections. 
Now the smaller corrections of the Higgs boson are being constrained, and 
there are tantalizing hints that the first direct evidence for a Higgs boson
has been seen at LEP. The mass of this Higgs boson is in agreement with 
the predictions from the precision measurements and is 
in a range that is consistent with supersymmetry.

This strategy of testing the consistency of theories will continue 
after a future discovery of supersymmetry and the required Higgs sector.
The purpose of the present note is to emphasize that in an era following
the discovery of a supersymmetric Higgs sector, there are some 
sets of observables for which precision measurements will be particularly 
powerful. In the minimal supersymmetric standard model (MSSM), 
supersymmetry together with gauge invariance impose 
constraints on the Higgs sector that gives
rise to mass sum rules. The Higgs sector of the MSSM contains three neutral 
Higgs bosons, $h$, $H$, and $A$ as well as two charged Higgs bosons, $H^\pm$.
The sum rules relate certain combinations
of mass-squares of the Higgs and gauge boson masses. 
The gauge boson ($W$ and $Z$) masses are now known very precisely. 
A future muon collider can produce neutral Higgs bosons in the 
$s$-channel\cite{Barger:1997jm,Barger:1995hr}.
By adjusting the energy of the machine so that one is sitting
on the Higgs boson resonances, the muon 
collider can produce thousands of Higgs bosons per year, 
and the mass and total 
width determined very precisely. Hence it will 
be possible to do precision tests of the sum rules. 

{\bf Higgs Boson Mass Sum Rule:\ }
At tree-level, the mass sum rule for the neutral states of the MSSM 
is\cite{Inoue:1982ej}
\begin{eqnarray}
&&M_h^2+M_H^2=M_A^2+M_Z^2\;,
\label{tree}
\end{eqnarray}
This is a {\it natural relation} in that it is satisfied at tree-level 
without a tuning of parameters. At tree-level it can be shown that 
$M_H\geq M_A$ so that one has the constraint $M_h\leq M_Z$. 
The sum rule does not depend
on any parameters like mixing angles or couplings; only the physical Higgs 
boson masses need to be measured to test the sum rule at the tree-level.
The sum rule receives a non-zero but finite and 
calculable correction from loop diagrams. The correction can be 
summarized as a contribution $\Delta$, so that
\begin{eqnarray}
&&M_h^2+M_H^2=M_A^2+M_Z^2+\Delta \;.
\label{sum}
\end{eqnarray}
One can solve this equation for the difference in the heavy Higgs boson
masses,
\begin{eqnarray}
&&M_H-M_A={{M_Z^2-M_h^2+\Delta}\over {M_A+M_H}}\;.
\end{eqnarray}
This form is instructive as it is clear that in the decoupling limit, 
$M_A, M_H\to \infty$, the mass difference $M_H-M_A$ becomes small. 
The mass difference is positive for most of supersymmetric parameter space, 
but it can take either sign depending on the details of 
the spectrum and couplings of the supersymmetric particles.
There are theoretical reasons to believe that the absolute value of 
this mass difference is small.
In the MSSM, large $M_A$ and large $\tan \beta$ give highly degenerate
heavy Higgs states separated by a few GeV or less. The ever increasing lower
bound on Higgs masses from the LEP experiments is gradually increasing the 
lower bound on $\tan \beta$ that is allowed in the MSSM making it more likely
that $\tan \beta$ is large. 

The leading contribution to $\Delta$ was first calculated
in Ref.~\cite{Berger:1990hg} and is 
\begin{eqnarray}
&&\Delta = {{3g^2m_t^4}\over {16\pi ^2M_W^2\sin ^2\beta}}\log 
{{m_{\tilde{t}_1}^2m_{\tilde{t}_2}^2}\over {m_t^4}}\;,
\label{leading}
\end{eqnarray}
where $\tilde{t}_1$ and $\tilde{t}_2$ are the top squark mass eigenstates.
There are smaller corrections from diagrams involving the lighter quarks,
gauge bosons, and their superpartners, and there are corrections from two
and higher loops. Following the renormalization of 
the sum rule, the radiative corrections to the light Higgs boson $h$
were isolated\cite{Okada:1991vk,Ellis:1991nz,Haber:1991aw}, 
and the tree level upper bound, $M_h\leq M_Z$, was
shown to no longer be satisfied. In fact, for most of parameter space, 
$\Delta$ contributes largely to the renormalization of the lightest Higgs ($h$)
mass for fixed $M_A$. Therefore a measurement of $M_h$ 
will provide the first test of radiative corrections in Higgs sector the MSSM.
A subsequent measurement of $\Delta$ as described below would constitute a 
precision test of these radiative corrections.

It should be emphasized that a precise measurement of $\Delta$ does not 
isolate any single supersymmetric mass or parameter, but rather
picks out a slice of parameter space. If the leading correction shown in 
Eq.~\ref{leading} was the only contribution, then a measurement of $\Delta$ 
would provide a measurement of the quantity 
$m_{\tilde{t}_1}^2m_{\tilde{t}_2}^2$ since the other quantities have already
been experimentally measured. The value of $\Delta$ can be calculated
theoretically for any choice of parameters and compared to the measured value.
The size of $\Delta$ is generally 
of order a 
few times $10^4$~GeV$^2$. The theoretical calculations for the radiative 
corrections have reached a high level of sophistication; see 
Ref.~\cite{Carena:2000dp,Espinosa:2000zm,Espinosa:2000df} 
and references therein for the present status.
One of the advantages of the sum rule is that the tree-level relation does 
not involve the supersymmetry parameter $\tan \beta$ which enters into the 
particle couplings. Any dependence on $\tan \beta$ enters only in the 
radiative correction $\Delta$, so the precision measurement discussed in 
this note can be carried out solely by measuring Higgs boson masses very 
precisely.

{\bf Precision Test of a Supersymmetric Higgs Sector:\ }
In the context of the MSSM, the correction $\Delta$ arises exclusively from
loop diagrams involving all the particles that couple to the Higgs bosons.
But in a more general model, the correction $\Delta$ might involve corrections
from some heavier Higgs boson states. So an experimental test of the sum 
rules probes radiative corrections in the MSSM, and probes for the presence
of heavier undetected Higgs bosons. As a concrete example, consider a 
multi-Higgs doublet supersymmetric model. Then the sum rule is generalized 
to be 
\begin{eqnarray}
&&\sum_{\rm CP-even} M_{H_i}^2=\sum_{\rm CP-odd} M_{A_i}^2+M_Z^2+\Delta \;.
\end{eqnarray}
where $M_{H_i}$ and $M_{A_i}$ represent the masses of CP-even and CP-odd 
Higgs bosons respectively. In this model where there are $2N$ Higgs doublets, 
there are $2N$ CP-even
mass eigenstates and there are $2N-1$ CP-odd mass eigenstates.
The mass difference between the lightest CP-odd Higgs boson and the 
second-lightest CP-even Higgs boson gets contributions not only from the 
radiative correction $\Delta$ but also from possibly 
small mass-squared differences
in the heavier Higgs boson states, $M_{H_i}^2-M_{A_i}^2$.

The presence of other electroweak representations of Higgs bosons can also
contribute an effective contribution to $\Delta$. For example, a small 
amount of mixing with a singlet Higgs boson will add a
contribution\cite{Drees:1989fc} that can be detected by accurately 
measuring $\Delta$.
The most important feature of supersymmetric models with Higgs sectors more
complicated than the MSSM is that the modifications to the mass sum rule
in Eq.~(\ref{tree}) appear already at the tree level.

In addition to the neutral Higgs boson mass sum rule in Eq.~(\ref{sum}), 
there is a sum rule
involving the charged Higgs boson,
\begin{eqnarray}
&&M_{H^\pm}^2=M_A^2+M_W^2+\tilde{\Delta} \;,
\end{eqnarray}
where $\tilde{\Delta}$ is the calculable correction to the tree-level sum
rule. The measurement of the radiative correction $\tilde{\Delta}$ is not
as interesting as the measurement of $\Delta$ we are highlighting in this 
note, since the mass of the charged Higgs boson can not be measured in 
$s$-channel production. However, a precise measurement of $M_{H^\pm}$ 
by other means might
prove useful as another probe of radiative corrections to the Higgs sector.

It has been shown\cite{Berger:1990hg,Gunion:1989pc,Gunion:1989dp}
that the loop contributions $\Delta $ and 
$\tilde{\Delta}$ are given exclusively by self-energy diagrams.
All contributions involving the loop corrections to the Higgs sector
mixing angles ($\alpha$ and $\beta$) conveniently
cancel out in the radiative corrections, so that measuring the couplings
is not necessary to obtain the experimental inputs 
to the highest order part (tree-level) of the sum rule. 

{\bf CP-violation:\ }
CP-violation can also be probed just by accurately measuring the Higgs 
boson masses. Loop-induced CP-violation can 
mix\cite{Pilaftsis:1997dr,Pilaftsis:1998dd,Pilaftsis:1998pe} 
the heavy Higgs CP-eigenstates,
$H$ and $A$, which generally leads to a shift in the relative positions 
of the mass-eigenstates (which are no longer the same as the CP-eigenstates). 
Higgs bosons that are highly degenerate in the absence
of CP-violation can be split when a CP-violating phase is 
nonzero\cite{Pilaftsis:1998dd,Demir:1999hj,Pilaftsis:1999qt}.
Since this splitting can be greater than a GeV, 
this constitutes another very interesting physical
effect that can be probed by accurately measuring $\Delta$.
On the other hand 
the mass splitting is a single parameter and ultimately one would want 
to exploit 
beam polarization to obtain more information.
If one has polarized beams available at the muon collider, then there are 
many more observables\cite{Asakawa:2001es}
that can be exploited to separate  
and measure the CP-mixing.
In fact, even if CP is conserved in the Higgs sector and the heavy Higgs
bosons are highly degenerate, one can use polarization
of the muon beams to separate the two resonances\cite{Grzadkowski:2000hm}.

{\bf An Example:\ }
In this section we present an example of the level of precision for the 
mass and total width of the heavy Higgs bosons that can be achieved 
through $s$-channel production.
We take as an example the following parameters: $M_A=350$~GeV, $\tan \beta=5$,
and take all supersymmetry breaking mass and mixing parameters to be 
1~TeV, e.g. $m_{q_{L,R}}^{}=A_t=A_b=1$~TeV. 
We also assume that CP-mixing 
between the heavy Higgs bosons is negligible. We use the program 
HDECAY\cite{Djouadi:1998yw} to calculate the radiatively corrected masses, 
decay widths, and branching ratios of the Higgs bosons. While it has been 
demonstrated that a muon collider is the optimal place to measure the 
light Higgs boson mass, $M_h$\cite{Barger:1997jm,Barger:1995hr},
a muon collider can also measure the heavy Higgs boson masses, $M_A$ and $M_H$,
very well in the $s$-channel production process. In fact, the muon collider 
may be the only 
possible machine that can separate two highly degenerate heavy Higgs and 
measure the mass difference, $M_H-M_A$.

\begin{figure}[tbp]
\postscript{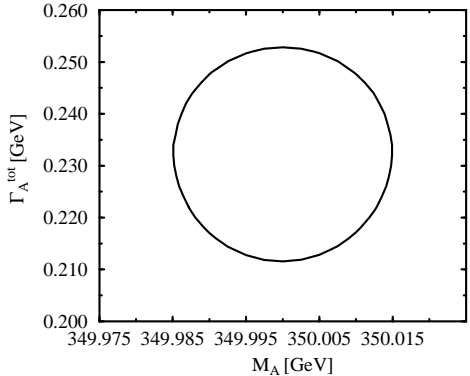}{0.77}
\caption{Higgs mass $M_A$ and total width $\Gamma ^{\rm tot}_A$ determination
for a scan over the resonance. 
The ellipse corresponds to the $\Delta \chi^2=1$ contour.}
\label{ellipse}
\end{figure}

We consider the process $\mu^+\mu^-\to A,H\to b\overline{b}$.
A scan over the Higgs resonances devoting $0.01$~fb$^{-1}$ of integrated 
luminosity to a sequence of center-of-mass energy values 
is employed to determine the
Higgs boson masses and total widths. The measurement is a counting experiment 
and does not require a precise energy determination of the $b$ jets; rather
at a muon collider the energy of the beams is expected to be known very 
well\cite{Raja:1998ip}
The result of such a scan is shown in 
Fig.~(\ref{ellipse}) for 11 scan energies separated by 0.1~GeV  around the 
Higgs resonance. We have assumed a Gaussian energy spectrum of each 
muon beam with an rms deviation $R=0.01\%$, and that the $b$'s are tagged
with a 50\% efficiency. One can simply multiply the 
ellipse in Fig.~(\ref{ellipse}) by an overall factor if one assumes a 
different tagging efficiency. The partial widths have not been allowed 
to vary in this scan, but relaxing this condition does not substantially
change the accuracy with which the Higgs boson mass can be determined.

One sees from Fig.~(\ref{ellipse})
in particular that one can determine the Higgs boson mass
with a $1\sigma$ error of just 15~MeV. The $1\sigma$ error on the width 
is roughly 20~MeV, which is about a 10\% measurement. A very 
similar determination can be attained for the $H$ boson since its 
total width and couplings to 
$\mu^+\mu^-$ are similar to those of the $A$ boson. 
The determined error generally shrinks as one goes to larger values of 
$\tan \beta$ as the couplings of the Higgs to $\mu^+\mu^-$ increases.
The Higgs boson widths also increase as $\tan \beta $ increases, and ultimately
the Higgs resonances overlap; when this happens, determining the mass
difference is problematic (see below).

Most of the discriminating power occurs for the luminosity devoted in the 
interval $M_A-\Gamma _A < \sqrt{s} < M_A+\Gamma _A$, but how much luminosity 
must be wasted on scan points outside this range depends on how well the 
Higgs boson mass is known prior to the scan. 
The masses of the heavy Higgs bosons must be known to less than 
or about 1~GeV before this type of scanning can be done, since it must 
be guaranteed that the Higgs peak cross section is within the scan energy 
range. Strategies for obtaining this precision have been discussed 
previously\cite{Barger:1997jm} and could take place at a future linear collider
or at a higher energy muon collider. The scenario in which things play out is
not known, so it is not clear how well the heavy Higgs boson masses will 
be known prior to the scan.
The light Higgs boson mass does give
us some information on the radiative corrections, and if some of the radiative
correction parameters were known (by a priori discovery of supersymmetry
and measurement of the particle masses and couplings), one could obtain a rough
indirect measurement of $M_A$ to 20\%\cite{Autin:1999ci}.
One can also discover the heavy Higgs bosons directly in the bremsstrahlung 
tail\cite{Barger:1997jm} at a muon or linear collider operating at an energy
above the Higgs masses. These rough determinations of the heavy Higgs boson 
masses could be followed by a rough scan that could pin down the mass(es) of 
$H$ or $A$ to a GeV.

The example in this section
shows that the muon collider with a reasonable amount of 
integrated luminosity can measure the heavy Higgs bosons of 
supersymmetric models to tens of MeV. This represents an extraordinary
probe of radiative corrections in the Higgs sector. The expected measurement
of the mass of the light Higgs $h$ is of the order of 100's of keV and is 
limited by the precision (for the expected integrated luminosity 
of 0.2~fb$^{-1}$) with which the beam energy can be measured through the 
spin precession of muons around the ring\cite{Raja:1998ip}. The $Z$-boson 
mass is currently known with an error of $2.2$~MeV from the LEP 
experiments\cite{Groom:2000in}.
So the dominant error
on the measurement of $\Delta$ will come from the errors on the mass 
measurements of the heavy Higgs bosons, $H$ and $A$. In the example, the 
contributions to the error on $\Delta $ are
\begin{eqnarray}
&&\delta (M_H^2)\sim \delta (M_A^2) \sim 10~{\rm GeV}^2\;.
\end{eqnarray}
This then results in a measurement of the radiative correction $\Delta$ of the
order of one part in $10^3$.

\begin{figure}[tbp]
\postscript{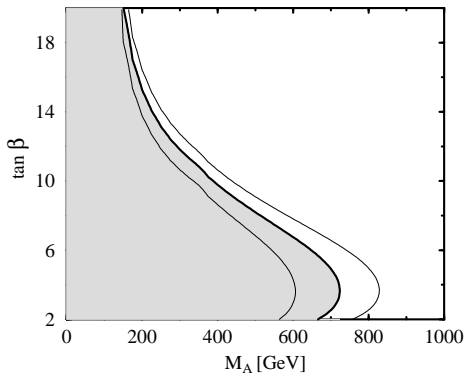}{0.77}
\caption{Region (shaded) for which the Higgs mass difference $|M_H-M_A|$
is sufficiently large ($>(\Gamma ^{\rm tot}_A+\Gamma ^{\rm tot}_H)/3$) 
that a scan 
over the $H$ and $A$ boson resonances can measure the two masses. Also shown 
are the $|M_H-M_A|=(\Gamma ^{\rm tot}_A+\Gamma ^{\rm tot}_H)/2)$ (leftmost 
contour) and $|M_H-M_A|=(\Gamma ^{\rm tot}_A+\Gamma ^{\rm tot}_H)/4)$ 
(rightmost contour) for comparison.}
\label{massvswidth}
\end{figure}

{\bf Higgs Boson Mass Degeneracy:\ }
When the heavy Higgs bosons become very degenerate, it will be much 
harder to determine the mass difference $M_A-M_H$. A rough rule for when 
the scan as described in this note will fail to be adequate at resolving 
the two Higgs resonances occurs for mass differences less than the one third
of the sum of the total widths of the heavy Higgs bosons, i.e.
\begin{eqnarray}
&&\left | M_H-M_A \right | < {1\over 3}\left (\Gamma ^{\rm tot}_A
+\Gamma ^{\rm tot}_H\right )\;.
\label{condition}
\end{eqnarray}
This condition assumes that the rms deviation $R$ of the energy spectrum of
each muon beam is sufficiently small so as to not smear two peaks together.
It is adequate to have $R=0.01\%$ for these heavy bosons, but a larger 
value such as $R=0.06\%$ would smear two partially overlapping resonances
together\cite{Barger:1997jm}. In cases where the Higgs bosons are 
sufficiently separated in mass, 
the rms deviation can be increased with a resulting
increase in luminosity because the heavy Higgs bosons $H$ and $A$
are very much broader than the light Higgs boson $h$ (a light 
Standard Model-like Higgs might require $R$ less than 0.01\% to fully exploit
the very narrow resonance).
The condition in Eq.~(\ref{condition})  
occurs typically for the larger values of both $M_A$ and $\tan \beta$.
Figure~(\ref{massvswidth})
shows this region for a particular choice of supersymmetric 
parameters ($M_{\tilde{q}}=A_q=1$~TeV). 
It should be kept in mind that different values for the 
supersymmetric masses and mixing, or the presence of CP-violation,
can produce heavy Higgs bosons that are shifted relative to each other, 
and would qualitatively change the contours in Fig.~(\ref{massvswidth}).

Even when one cannot experimentally distinguish the overlapping heavy Higgs 
bosons, one can still derive an upper bound on their mass difference if one
makes the hypothesis that the one resonance peak that is being observed 
is two overlapping Higgs bosons. Furthermore, techniques exploiting any 
possible polarization of the muon collider can by used to unravel the 
CP-even $H$ boson from the CP-odd $A$ 
boson\cite{Asakawa:2001es,Grzadkowski:2000hm}.

{\bf Summary:\ }
We have demonstrated that the study of the heavy Higgs bosons of the MSSM in 
the $s$-channel at a future muon collider can be combined with the mass 
measurement of the light Higgs boson to sensitively probe radiative corrections
to the MSSM Higgs sector. Very accurately measuring the mass difference
of the heavy neutral Higgs bosons of the MSSM can probe possible 
CP-violation or nonminimality of the Higgs sector. 
Comparison of the experimentally measured Higgs 
boson masses with calculations of the virtual effects of Standard Model 
and supersymmetric particles can give a precise test of the MSSM or a 
definite prediction that must be satisfied by the supersymmetric spectrum.
An important question is whether the theoretical calculations will progress
far enough to make full use of the possible experimental determination of the
radiative corrections as suggested in this note.
This would probably require the calculation of the subleading two-loop 
contributions of ${\cal O}(M_Z^2\alpha _t\alpha_s)$ and 
${\cal O}(M_Z\alpha_t^2)$ as well as calculations of the self-energy diagrams
without making the zero-momentum approximation on the external legs.
  

\vspace{0.5cm}


This work was supported in part by the U.S.
Department of Energy
under Grant No. 
No.~DE-FG02-91ER40661.




\begin{references}

\vspace{-1cm}

\bibitem{Barger:1997jm}
V.~Barger, M.~S.~Berger, J.~F.~Gunion and T.~Han,
Phys.\ Rept.\ {\bf 286}, 1 (1997)
[hep-ph/9602415].

\bibitem{Barger:1995hr}
V.~Barger, M.~S.~Berger, J.~F.~Gunion and T.~Han,
Phys.\ Rev.\ Lett.\ {\bf 75}, 1462 (1995)
[hep-ph/9504330].

\bibitem{Inoue:1982ej}
K.~Inoue, A.~Kakuto, H.~Komatsu and S.~Takeshita,
Prog.\ Theor.\ Phys.\  {\bf 67}, 1889 (1982).

\bibitem{Berger:1990hg}
M.~S.~Berger,
Phys.\ Rev.\ D {\bf 41}, 225 (1990).

\bibitem{Okada:1991vk}
Y.~Okada, M.~Yamaguchi and T.~Yanagida,
Prog.\ Theor.\ Phys.\  {\bf 85}, 1 (1991).

\bibitem{Ellis:1991nz}
J.~Ellis, G.~Ridolfi and F.~Zwirner,
Phys.\ Lett.\ B {\bf 257}, 83 (1991).

\bibitem{Haber:1991aw}
H.~E.~Haber and R.~Hempfling,
Phys.\ Rev.\ Lett.\  {\bf 66}, 1815 (1991).

\bibitem{Carena:2000dp}
M.~Carena, H.~E.~Haber, S.~Heinemeyer, W.~Hollik, C.~E.~Wagner and G.~Weiglein,
Nucl.\ Phys.\ B {\bf 580}, 29 (2000)
[hep-ph/0001002].

\bibitem{Espinosa:2000zm}
J.~R.~Espinosa and R.~Zhang,
JHEP {\bf 0003}, 026 (2000)
[hep-ph/9912236].

\bibitem{Espinosa:2000df}
J.~R.~Espinosa and R.~Zhang,
Nucl.\ Phys.\ B {\bf 586}, 3 (2000)
[hep-ph/0003246].

\bibitem{Drees:1989fc}
M.~Drees,
Int.\ J.\ Mod.\ Phys.\ A {\bf 4}, 3635 (1989).

\bibitem{Gunion:1989pc}
J.~F.~Gunion and A.~Turski,
Phys.\ Rev.\ D {\bf 39}, 2701 (1989).

\bibitem{Gunion:1989dp}
J.~F.~Gunion and A.~Turski,
Phys.\ Rev.\ D {\bf 40}, 2333 (1989).

\bibitem{Pilaftsis:1997dr}
A.~Pilaftsis,
Nucl.\ Phys.\ B {\bf 504}, 61 (1997)
[hep-ph/9702393].

\bibitem{Pilaftsis:1998dd}
A.~Pilaftsis,
Phys.\ Lett.\ B {\bf 435}, 88 (1998)
[hep-ph/9805373].

\bibitem{Pilaftsis:1998pe}
A.~Pilaftsis,
Phys.\ Rev.\ D {\bf 58}, 096010 (1998)
[hep-ph/9803297].

\bibitem{Demir:1999hj}
D.~A.~Demir,
Phys.\ Rev.\ D {\bf 60}, 055006 (1999)
[hep-ph/9901389].

\bibitem{Pilaftsis:1999qt}
A.~Pilaftsis and C.~E.~Wagner,
Nucl.\ Phys.\ B {\bf 553}, 3 (1999)
[hep-ph/9902371].

\bibitem{Asakawa:2001es}
E.~Asakawa, S.~Y.~Choi and J.~S.~Lee,
Phys.\ Rev.\ D {\bf 63}, 015012 (2001)
[hep-ph/0005118].

\bibitem{Grzadkowski:2000hm}
B.~Grzadkowski, J.~F.~Gunion and J.~Pliszka,
Nucl.\ Phys.\ B {\bf 583}, 49 (2000)
[hep-ph/0003091].

\bibitem{Djouadi:1998yw}
A.~Djouadi, J.~Kalinowski and M.~Spira,
Comput.\ Phys.\ Commun.\  {\bf 108}, 56 (1998)
[hep-ph/9704448].

\bibitem{Raja:1998ip}
R.~Raja and A.~Tollestrup,
Phys.\ Rev.\ D {\bf 58}, 013005 (1998)
[hep-ex/9801004].

\bibitem{Autin:1999ci}
B.~Autin, A.~Blondel and J.~Ellis,
CERN-99-02.

\bibitem{Groom:2000in}
D.~E.~Groom {\it et al.}  [Particle Data Group Collaboration],
Eur.\ Phys.\ J.\ C {\bf 15}, 1 (2000).

\end{references}
\end{document}